# EXAMINATION OF THE CORRELATION BETWEEN WORKING TIME REDUCTION AND EMPLOYMENT

Virginia Tsoukatou

Department of Political Science and International Relation, University of Peloponnese, Greece

**Abstract**

In recent years, it has been debated whether a reduction in working hours would be a viable solution to tackle the unemployment caused by technological change. The improvement of existing production technology is gradually being seen to reduce labor demand. Although this debate has been at the forefront for many decades, the high and persistent unemployment encountered in the European Union has renewed interest in implementing this policy in order to increase employment. According to advocates of reducing working hours, this policy will increase the number of workers needed during the production process, increasing employment. However, the contradiction expressed by advocates of working time reduction is that the increase in labor costs will lead to a reduction in business activity and ultimately to a reduction in demand for human resources. In this article, we will attempt to answer the question of whether reducing working hours is a way of countering the potential decline in employment due to technological change. In order to answer this question, the aforementioned conflicting views will be examined. As we will see during our statistical examination of the existing empirical studies, the reduction of working time doesn't lead to increased employment and cannot be seen as a solution to the long-lasting unemployment.

## INTRODUCTION

In recent years, it has been debated whether a reduction in working hours would be a viable solution to tackle the unemployment caused by technological change. The improvement of existing production technology is gradually being seen to reduce labor demand. Although this debate has been at the forefront for many decades, the high and persistent unemployment encountered in the European Union has renewed interest in implementing this policy in order to increase employment. According to advocates of reducing working hours, this policy will increase the number of workers needed during the production process, increasing employment. However, the contradiction expressed by advocates of working time reduction is that the increase in labor costs will lead to a reduction in business activity and ultimately to a reduction in demand for human resources. In this article, we will attempt to answer the question of whether reducing working hours is a way of countering the potential decline in employment due to technological change. In order to answer this question, the aforementioned conflicting views will be examined. Furthermore, we will statistically examine numerous empirical studies in order to make some conclusions regarding the impact of working time reduction to the employment levels.

## ADDRESSING THE CONSEQUENCES OF THE REDUCTION OF WORKING TIME IN EMPLOYMENT AND PRODUCTIVITY

In this section, three economic models will be examined: the competitive market model, the monopsony model, and the collective bargaining model. The competitive market model is a useful starting point in looking at the impact of reducing working hours. As we will see, compulsory reductions in working time do not improve social welfare or increase employment. Consequently, the implementation of the measure under consideration is not a way of countering the potential contraction of employment due to technological change. However, since there is rarely a perfectly competitive market, in reality, other alternative models are then considered to form a more complete view of the issue. In conclusion, we could observe that although the reduction of working hours cannot be considered as a panacea since in most cases it has a negative impact on the economy, nevertheless under very specific conditions under consideration it could raise the level of employment.

### Competitive market theory

*Correlation between working time and level of employment*

The main argument used by proponents of working time reduction is that the amount of production of products and services in an economy is a given, so a possible reduction in working time will lead to a redistribution of a certain amount of work to more workers, increasing employment. In a perfectly competitive economy, any mandatory reduction in working hours would create flaws and limitations in a situation where the theoretical model allocates resources efficiently (Cahuc & Zylberberg, 2008). According to the theory, a reduction in working hours could not lead to an increase in employment. When the maximum working time is reduced legally, companies have three options: to rely on overtime, to hire staff for limited working hours or to reduce production levels. If we consider for the sake of analysis that in this model firms want to maintain stable production and that the only relevant labor cost is the hourly wage of workers then the problem of reducing working hours can be completely solved by recruiting new employees for the hours left. In this case, reducing the working week will lead to an increase in employment. Even under these conditions, however, any reduction in unemployment will increase the demand for labor and thus lead to an increase in average wages. As a result, it will gradually reduce the employment level again (Plantenga & Dur, 1998). However, it would be methodologically incorrect to ignore all the other costs associated with hiring new employees (costs associated with their recruitment and training, labor contributions that are in many cases unrelated to working hours, etc.). The aforementioned costs differentiate labor costs by making a new employee's hourly labor cost higher than an existing employee. If we take into account that overtime is usually costed higher than the conventional working time we realize that in this case too, unit costs are rising and as a result, the increase in employment is rather doubtful. Consequently, if a company decides to use overtime, a reduction in the working week may not increase the marginal cost for each hour of overtime but increases the marginal cost of hiring a new employee. So businesses prefer to "buy" more hours of work from existing employees but not hire new ones. In this way, when the working time is reduced, the demand for overtime is increased but this is not transformed into an increase in employment and a reduction in unemployment. Accordingly, some companies may prefer to modify their production process (purchase of new capital equipment that will replace the labor force) or their output level as a consequence of reduced working hours for reasons of sustainability. In this case, as production costs increase (since a given labor cost is split into fewer hours of work) firms may choose to produce less, reducing not only the hours worked but also the number of employees.

If we take into account that the increase in labor costs can cause workers to replace capital equipment, then we realize that not only is the reduction in working hours not a response to the decline in employment due to technological change but, on the contrary, may accelerate the pace with which companies decide to upgrade their production technology (Börsch-Supan, 2002).

*Correlation between working time and productivity*

In business, the unit cost of the product is important, not the absolute wage cost per se. By extension, there are other parameters too in the production function that determine the impact of reducing weekly working hours on employment at a given level of production. One of these parameters is the productivity of workers in relation to working hours, since, for example, when the working days are longer, more fatigue occurs and their marginal productivity decreases. Productivity is expressed as the number of product units produced at a given time using specific resources (in this case workers). In conclusion, the more productive an economy is, the more product is produced in less time. Productivity is therefore influenced not only by production technology but also by the quality of the workforce (how well employees are, what level of education they have, how much they strive to get their job done, etc.) (De Spiegelaere & Piasna, 2017).

The theoretical approach that underpins productivity growth as a result of reduced working hours attributes it to three factors: biological, organizational, and employee motivation factors. The first factor (biological) is related to the fact that fewer work hours reduce the level of fatigue of employees. At the same time, less fatigue improves concentration resulting in increased productivity. The second factor relates to the improvement of the production process resulting from the extended hours of use of capital equipment and the increased innovation and creativity that allows employees to carry out more work in less time. The increase in capital productivity, in this case, may be associated with an increase in the hours the production line operates. Assuming, for example, that the company previously operated two shifts of 8 hours each (16 in total) and after the reduction of working hours 3 shifts of 6 hours were created, we observe that the operating cost of capital equipment is reduced as operating and maintenance cost data are distributed to more hours of use and therefore more production volume. Through this expanded use of production equipment, there will be a reduction in unit production costs. This could reduce working hours while maintaining the level of pre-work time levels. However, this would be applicable to companies engaged in product production while it would be doubtful whether the same benefits would exist for service providers unless there was a corresponding increase in demand for services (Bosch & Lehndorff, 2001; Hoel & Vale, 1986). Finally, a third factor that is difficult to quantify, however, relates to employees' motivation to work more efficiently, even sacrificing their break time, for example in response to less work time.

*Prerequisites for increasing employment in an environment of reduced working hours*

As stated above, according to the theory that employment could potentially increase due to reduced working time. In any case, however, there is a link between working hours and employee earnings, which casts doubt on the acceptance of the measure of work week restriction by labor unions and employees (Cahuc & Zylberberg, 2008). When the limitation of working time is combined with increases in hourly earnings so that the monthly wage is not reduced, then there is essentially a double unequal redistribution of available resources, and this indirect increase in pay will be even greater if typical annual pay increases are taken into account.

In this case, the reduction of working hours creates a clear increase in production costs. In order to be viable to reduce the workweek and ultimately to increase employment, there must either be a reduction in remuneration equal to a reduction in working hours or an overall wage bargain where lower annual increases will be agreed to compensate for the working time reduction. If in this way, the impact of the increase in wage costs is minimized by productivity gains, then unit production costs will remain stable thereby improving the performance of the economy in employment. In addition, another way that could the working time reduction has a positive effect on employment is to disconnect the time of using capital equipment from the basic business hours. In production plants that make extensive use of production machinery, most of the cost is related to the limited time of use of the machinery and not to the wage cost. If a possible reduction in working time is accompanied by negotiation and ultimately agreement, to limit working hours and steady earnings in return for extended line operation through shifts even in non-social hours (for example late at night) then it will be achieved a reduction of operating expenses. Increasing use of capital equipment can produce the same number of product units in a shorter period of time, reducing the hours that production machinery is not used (Bosch & Lehndorff, 2001).

## Theory of collective bargaining

As mentioned in the previous section, the impact of reducing working hours largely depends on the 'reaction' of wage costs to this reduction in working time. Therefore, it is of particular interest to look at the balances that are formed when this restriction occurs in economies with extensive use of collective bargaining. According to Cahuc et al., (2008), the effect of reduced working time depends on the bargaining power of labor unions and employers, current legislation, workers' preferences for reduced working hours, and the degree of coordination encountered in the economy. Cahuc et al., (2008) in their theoretical model consider the case of an economy where working hours and earnings are determined after collective bargaining between employers and labor unions. In the economy in which these partners operate, there is a statutory upper limit on working time which is separate from normal working hours. The difference between the working time provided for in each employee's contract and the ceiling shall be remunerated as overtime with a higher hourly wage than usual. The main objective of the trade unions is to reach the best possible agreement for their members which provides as few hours as possible and higher wages. The result of collective bargaining with businesses is what will determine working hours, earnings, and ultimately the number of jobs. According to the predictions of this theoretical model, the stronger the trade unions and the better their bargaining position, the shorter the working hours agreed with employers.

In addition, a firm's negotiating position on the issue of working hours is related to the firm's market position and the degree of centralization of collective bargaining. When the company is a market leader (thus facing less competition) and at the same time collective bargaining is highly centralized, it is more likely to reach an agreement that provides for fewer hours of work. Correspondingly, when labor unions emphasize maintaining a level of employment, they tend to agree to more working hours in order to divide wages into longer working time to keep wage costs lower. In this case, even higher levels of employment could be achieved through wage moderation and the high degree of coordination observed.

In particular, according to Cahuc et al. (2008), the elasticity of wage hours relative to working time (H) depends on the number of hours worked. If elasticity is a positive number, it is concluded that any reduction in working hours will result in a reduction in the employee's salary. In addition, the longer the working hours for an employee, the greater the reduction in his pay as a result of reduced working hours. Consequently, it would be easier to apply a reduction in working hours in cases where the working hours are sufficient (because the reduction in pay would not represent a disproportionately large share of pay) and workers attach great importance to more leisure time.

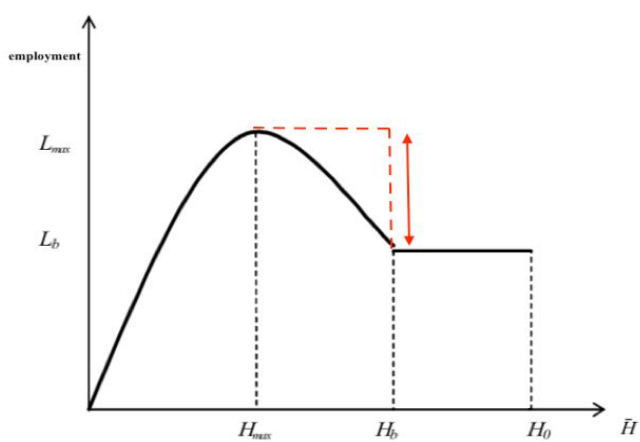

FIGURE 1. EMPLOYMENT IN A COLLECTIVE BARGAINING ENVIRONMENT
*Source*: Adapted from Cahuc et al., (2008)

Figure 1 illustrates the employment response to wage reductions in an environment of widespread collective bargaining and centralization. Hb is the number of working hours determined in the context of negotiations between employers and labor unions. For Hb working hours employment is at Lb level. After collective bargaining, the working hours at Hb level are agreed so that the individual working hours are equal to the parallel straight line to the right (Lb, Hb). As the graph shows, the maximum employment level for the economy is at Lmax. To achieve this level of employment the working hours must be Hmax. The reduction of working hours in the collective bargaining environment we are considering will only increase employment if employees are working longer hours than at Hmax because below that, the wage increases extensively as a result of reduced working hours (large elasticity) preventing the creation of more jobs. The Hmax point is what the trade unions want in a bargain and can be achieved when they use all their bargaining power. The distance between Hb and Hmax decreases as the market power of the company increases. Also may decrease when the preference for the leisure time of the employees is increased or when the bargaining power of the labor unions decreases. By extension, when all of the above factors are present at the same time, it would be likely that Hmax would be very close to Hb and that employment could be slightly increased through a mandatory reduction in working hours. In this sense, as we shall see, the results should be similar to those of monopsony.

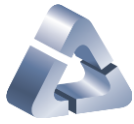

Reducing working time also has an impact on employee productivity as it relates to the performance and effort maid. In the case of centralized collective bargaining systems where labor unions represent a large proportion of the workforce, it is particularly likely to agree to reduce working hours for their members. In return, it may be agreed to adopt a wider working time that would allow businesses to operate at a time that previously didn't. Based on figure 1 we can see that the Hmax point could increase as the elasticity of the working hours increases. This elasticity is related to the possible extent of restructuring of the production structure. Consequently, it will be smaller in cases where businesses can reduce labor demand through the application of new production methods (Cahuc et al., 2008). In conclusion, the labor flexibility agreed by labor unions in order to achieve fewer working hours may lead to worse working conditions despite the reduced working time.

## Theory of monopsony power

In a market where businesses have monopsony power, it is observed that a small mandatory reduction in working hours will increase employment. Conversely, when the reduction in working hours is large, the impact on employment will be negative (Cahuc & Zylberberg, 2008; Marimon & Zilibotti, 2000). In a monopsony market that does not have an upper limit on working time, companies will choose the right combination of wages and working hours to maximize their profits. By extension, they will opt for a salary lower than the salary in the competitive market and working hours higher than those in the competitive market. Therefore, the level of employment in the economy is lower. If the state imposes a legal upper limit on working time, the monopsony market restrictions on working hours will be lifted and, consequently, the employment rate will increase as the benefits of workers from employment will increase. But while legislative intervention improves employment, alone is not enough to reach the optimum point. At the same time, the monopsony market should have a minimum wage in combination with a reduction in working time to improve social well-being.

Trying to graphically illustrate this view (Figure 2), we observe that the working hours in a monopsony are $H_M$. If these hours are more than the maximum working hours then they do not impose any restrictions on the monopoly firms that continue to employ $L_M$ labor for $H_M$ hours per day. If the working hours selected by the firms in the market are more than $H_M$ then the government restriction on working hours limits business policies and sets the employment level to $L_M$. If in this monopsony the working hours are set at a level H which is lower than the upper limit $H_M$ then employment is plotted along the horizontal straight line to the right of H.

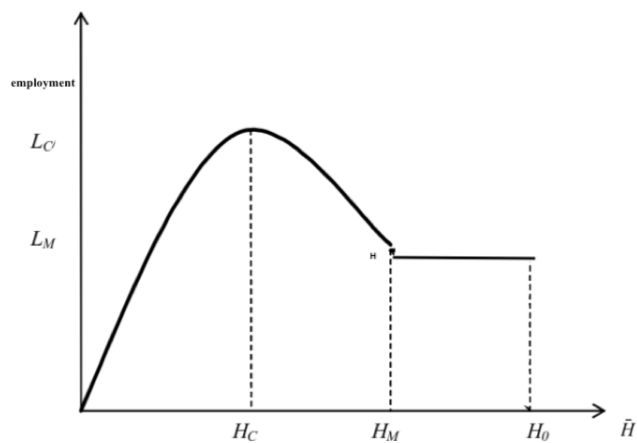

FIGURE 2. EMPLOYMENT IN A MONOPSONY
Source: Adapted from Cahuc et al., (2008)

Keeping the above in mind, we observe that there would be an increase in employment if the hours worked were more than the hours $H_C$ that are the hours worked by an employee in a competitive market. In this case, employment would reach $L_C$. Correspondingly, when working hours are reduced to below the $H_C$ level, the impact on employment is negative. In conclusion, employment reaches its highest level when the government exogenously sets an upper limit of employment equal to the $H_C$ point which constitutes the working time in a competitive economy. Even so, however, the level of employment is lower than that of the competitive economy due to lower remuneration in a monopsony (Cahuc et al., 2008).

In conclusion, the monopsony market model (although it is utopian and not present in everyday life) ,as well as collective bargaining, shows that a slight reduction in working hours improves employment temporarily but in the long run does not substantially affect the stabilization of the economy at a higher level of employment. However, employees have different priorities regarding their professional lives. Some prefer more leisure time (fewer working hours) while others prefer the higher financial welfare that most working hours give them. Given that employees have different productivity and working time priorities, they are unlikely to improve overall long-term employment and well-being by simply setting a ceiling on working time without taking into account the particularities of each employee and each business (Cahuc & Zylberberg, 2008). In addition, as we observe, the view that the level of the product remains constant and therefore a reduction in the basic working hours will in itself lead to an increase in employment is misleading. Accordingly, the impact of the reduction in working hours is determined by the working time that companies choose to 'buy' on the basis of the resulting labor costs. When this time is reduced compared to the previous working week, then there is a decrease in employment. By extension one of the most important conditions for maintaining or slightly increasing employment is wage restraint. As Cahuc and Zylberberg (2008) observe, a possible 10% reduction in working hours reduces profitability and output in each case examined unless accompanied by a corresponding 10% reduction in earnings. In cases where earnings have remained stable, a reduction in working time has called into question the viability of businesses. As can be seen from the above, when a reduction in working time is not accompanied by a corresponding reduction in earnings, it is rarely beneficial. Finally, we examined the case of bargaining over working time and wages through collective bargaining between labor unions and employers. To a large extent, the unions' attitude is related to their bargaining power and the economic environment in which they operate. In this case, too, it was observed that when it came to improving employment, labor unions tended to agree on wage restraint and reduced working hours in return for accepting work during 'non-social hours'.

By extension, although the competitive market model seems more realistic in its theoretical predictions, it would be useful to move on to the next sections in empirical research based on a meta-analytic technic that examins the impact of reduced working hours on employment and productivity in order to form a clearer picture of whether reducing working hours is a way of tackling the potential decline in employment due to technological change.

## AN EMPIRICAL EXAMINATION OF THE WORKING TIME REDUCTION IN THE NETHERLANDS AND FRANCE

As mentioned above, monopsony is an extreme form of an imperfectly competitive economy created in cases where combined labor mobility and high costs of entry into new markets are combined. In the current form of the economy, it is particularly difficult to come up with monopsonistic markets, so this approach is primarily a theoretical prediction of the impact of reduced working hours on employment but is not particularly likely to be implemented in practice. As an extension, in this section, we will look at the other two theoretical approaches mentioned: competitive market theory and collective bargaining theory. Examining how the reduction of working time in the Netherlands and France has evolved, we will attempt to illustrate the impact of working hours on an economy in which the reduction of working time is governed by the theory of collective bargaining and one that is governed by the theory of competitive market respectively.

### *Netherlands*

The reason we chose the Netherlands as a representative example of collective bargaining theory is the structure of the country's labor relations characterized by a high degree of coordination, high participation in labor unions and widespread use of collective bargaining. An interesting element, as we shall see below, is the acceptance by the workers of salary reduction in the form of non-compensation for hours not worked due to the limitation of working hours. This is because of the interesting point, as mentioned in the previous section, is the relation between unit labor costs and employment.

## Structure of industrial relations in the Netherlands

The reduction in working time has been the subject of economic analysis in the Netherlands for years. Proponents of this approach have argued that a reduction in working hours is a prerequisite for a significant reduction in the unemployment rate. Apparently, supporters of this view perceived labor demand as a given size. Consequently, a reduction in working hours would result in an increase in available jobs and as a result to a reduction in unemployment. In contrast, employers disagreed with the reduction of working time as they believed it would lead to increased hourly labor costs, lower production, lower employment levels and ultimately fewer jobs. In the late 1970s, the economy began to show signs of stagnation. Growth slowed and unemployment rose rapidly (from about 1-2% in the 1960s to 6% in the late 1970s (van Ours, 2006)). The protection of jobs was a priority for labor unions during this period. As a result, the reduction in working hours has been reintroduced to the public debate as a tool to maintain employment at previous levels. Employers, on their part, fearing that a further reduction in working hours would result in an increase in hourly labor costs, did not want any further reduction in working time. Alternatively, they offered workers early retirement plans (Plantenga & Dur, 1998). Gradually, as the economy has stagnated, part-time work is beginning to be seen as a way of restructuring the labor market. In the early 1980s, unemployment reached 12% (in 1984). In any case, labor unions saw employment protection as a clear priority. As a result, they were increasingly accepting flexible working relationships. In this context, employers and employees have agreed to reduce working time in return for wage restraint and the abolition of automatic wage adjustment (Wassenaar Agreement). At the same time, as a consequence of rising unemployment, flexible forms of employment were partially diminished as a result of state policies that improved the legal protection of part-time workers.

In 1998, through numerous collective agreements in different sectors, the working week was again reduced to 36 hours. Through the adoption of more flexible forms of employment, unemployment declined to 2% in 2001. In the following years, although there was some increase in unemployment, it nevertheless stabilized at an acceptable level (close to 5-5.5%). Following the Wassenaar Agreement, part-time workers enjoy the same social benefits as full-time workers, while being paid in proportion to their actual working time. The Wassenaar Agreement was essentially the turning point for the labor market as it left behind a long period of ideological differences between social actors and established a period of cooperation and harmony. It is precisely this coordination between the social partners and the consensus that has brought about such a dramatic improvement in the performance of the economy. The Dutch labor market is characterized by a high degree of cooperation between the social partners. In the context of the dialogue between employers and labor unions it was agreed that since the primary priority of workers was to protect employment, workers' demands should not increase labor costs or reduce the operating hours of the enterprise. It is therefore observed that in the case of the Netherlands wages did not remain stable on a monthly basis despite reduced working hours. The reduction in monthly earnings was equivalent to a decrease in working time. In conclusion, for its part, the state has committed and reformed the entire social security system. Now part-time workers are no different from full-time workers in the area of taxation and social security. Subsequently, in the context of this reform of the welfare state, replacement rate of unemployment benefits in relation to the last wage and the duration of their payment have been reformed.

## Evaluation of the Dutch model of reduction of working time

By looking at the Dutch example, we aim to examine whether reducing working hours is a way to deal with the possible contraction of employment due to technological change in a highly coordinated environment with high participation in labor unions and centralized collective bargaining. It is recalled that according to the predictions of the theory of collective bargaining when labor unions emphasize maintaining a level of employment tend to keep wage costs lower. Therefore, in some cases, unemployment could be reduced through wage moderation and the high degree of coordination observed. In the case of the Netherlands, all the above factors apply therefore we would expect a clear increase in productivity, a decrease in unemployment and an increase in employment.

Undoubtedly, the unemployment statistics are particularly important indicators of labor market performance. However, in many cases, they do not describe the full picture. In the case of the Netherlands, much of the reduction in unemployment is due to the systematic shift of workers out of the labor market (for example, through early retirement or disability benefits). Therefore, statistics examining the employment and not unemployment are of particular importance. In the Dutch labor market, non-employment rates are high despite falling sharply between 1985 and 2000. It is indicative that in 1985, 47.7% of productive citizens were out of the labor market due to unemployment, disability, education, early retirement or other personal reasons. This figure dropped significantly to 34.4% in 2000 but is still high (van Ours, 2006: 137). If we look at the rates of non-participation in the labor market for males for the same period, we will see that they have remained stable. Consequently, this decrease is mainly due to increased participation in the labor market for women. For many years, women's participation in the production process was limited in the Netherlands, compared to other European countries. By and large, their increase in labor market participation is due to the "catch up effect" and to a lesser extent to the increase in part-time employment. This is because when an economy starts at a lower level of employment for women compared to the rest of Europe and tries to approach the European average, the increase is actually higher than the rest of the countries but this is due to the lower level of employment for women initially. A second reason for low labor market participation is as mentioned above a large number of citizens receiving welfare support. According to the Dutch legal framework, one can receive a disability benefit after the expiry of the sickness benefit which is paid for one year. While sickness benefit is temporary and is intended to constitute an income safety net, the disability benefit is not temporary in nature and is intended for employees who are wholly or partially unable to receive a salary from their work (as a result of disability or other problems for whatever reason). The broad wording of the legal framework has led to the phenomenon of a large proportion of the working population receiving disability benefits. Specifically, in 1999, 17.9% of 55-65 year olds received a welfare allowance (van Ours, 2006). Under the law, for an employee to receive this benefit, there was no need for an objectively obvious medical reason. As a result, an increasing number of workers were receiving full or partial disability benefits for psychological reasons (Spithoven, 2002).

The legal framework mentioned above has become the vehicle by which a large proportion of workers near retirement age leave the labor market (artificially reducing the unemployment rate). The number of these workers did not appear in the official unemployment data but at the same time, a large number of former workers applied for and received disability benefits (CPB, 1998). Already in the late 1970s, the number of beneficiaries increased rapidly, with unemployment rising at a slower rate than the number of beneficiaries (Spithoven, 2002). In conclusion, in large part because of the legal protection of workers in the event of redundant dismissals, employers and employees preferred the choice of layoff benefits in order to reduce their staff, largely hiding much of the unemployment rate (Hassink, 1996 ch. 6). Table 1 shows the percentage of the workforce receiving full-time equivalent benefits for 1980-1999.

TABLE 1. PERCENTAGE OF THE WORKFORCE RECEIVING BENEFITS FROM 1980-1999

|  | 1980 | 1990 | 1999 |
|---|---|---|---|
| Belgium | 17.4 | 24.4 | 23.6 |
| Netherlands | 15.9 | 19.9 | 17.8 |
| Germany | 15.2 | 18.1 | 22.4 |
| France | 13.9 | 20.2 | 24.2 |
| Denmark | 20.1 | 23.2 | 23.1 |
| Sweden | 16.1 | 17.0 | 20.0 |
| UK | 15.2 | 18.5 | 18.9 |
| Spain | 8.3 | 12.3 | 11.2 |

*Source*: Adapted from Marx (2007)

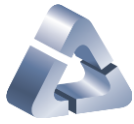 **Journal of Applied Economics and Business**

The percentage shown in the table is calculated as the ratio of the recipients of full-time equivalents to the total workforce. Consequently, although there is a downward trend, this could be attributed to an increase in the number of the working population in absolute terms and not to a real reduction in benefits recipients. Despite the strong job creation that could significantly reduce dependency on the welfare state, the number of people who are stopped being financially dependent on welfare benefits is very limited. In this sense, the Dutch model failed to bring these beneficiaries back into the labor market. Lastly, although the Dutch labor market has seen a large increase in the number of people employed, nevertheless taking into account the working hours per year, then this increase is not as impressive as it initially seemed. If we calculate this increase in full-time equivalents we will see that employment growth is 15.9% lower than when absolute jobs are calculated (Spithoven, 2002). Although the government expected to reduce unemployment by adopting more flexible forms of work, it remained relatively stable during 1987-1996 as most new part-time jobs were occupied not by registered unemployed but by new entrants (mainly students and former housewives) entered the labor market.

## Conclusions regarding the "Dutch Miracle"

The Dutch example was used to answer the research question of whether reducing working hours is a way of dealing with the possible contraction of employment due to technological change in a highly centralized and coordinated environment. As it turned out, the Netherlands was an excellent case for analysis because it meets the following conditions: particularly high centralization of the labor market with a high degree of coordination, the achievement of wage restraint through proportional wage reductions (due to reduced hours) and extensive reductions in working time. Even in an ideal model such as the one that closely resembles the model of collective bargaining, we see that the results are not what we would expect.

As has been said earlier, a reduction in working hours is not a panacea, and it is doubtful whether it could solve the contraction in employment as a result of technological change as it does not lead to increased productivity. In the area of employment and unemployment, we see that this reform has not "magically" reduced unemployment. What has happened is the decline in the natural rate of unemployment due to changes in the labor market. In the late 1990s and early 2000s, real unemployment fell below its normal rate, leading gradually to wage increases and as a self-fulfilling prophecy to rising unemployment again. With the structure of the labor market, this vicious cycle will be repeated over time based on the phase of the business cycle in which the economy is located. If we combine the above with the high degree of dependency on the welfare state and the low employment rates at older ages associated with a large number of early retirement financed by the welfare state we will realize that the reduction in time In the case of the Netherlands, it is clearly not a response to the decline in employment. In conclusion, the "Dutch miracle" consists of lower than expected unemployment reductions as reflected by full-time equivalents, high "hidden" unemployment, and low productivity growth.

## *France*

By looking at the French example, we aim to examine whether reducing working hours is a way to deal with the possible contraction of employment due to technological change in a competitive market environment. France is a country that is not particularly influenced by trade unions, nor is it highly coordinated. Participation in labor unions decreased from 22% in 1970 to 9% in 1992 (Hunt, 1998). To a large extent, these idiosyncratic differences are also reflected in the structure of labor relations, since negotiations on reducing working time are usually initiated by the workers rather than the state (Boulin, 1993). An interesting element, as we shall see below, is the different reaction of the labor market after the 1982 reforms and the Aubry reforms, demonstrating that the reduction of working time without subsidies by the State has a negative impact on employment.

## Structure of industrial relations in France

The issue of reducing working time has been debated for years. Working hours first began to decline significantly in the 1970s. The sharp decline in working time took place between 1974 (from 48 hours) and 1981 (to 40 hours) (Cahuc et al., 2008). The Mitterrand government in May 1982 completely unexpectedly reduced working hours again to 39 although there was no particular support for the measure by employers. At the same time, it amended overtime legislation. Under the new scheme, the first 4 hours of overtime were paid 25% more than the standard hourly wage. Employment beyond the first 4 hours was paid 50% higher hourly wages. Furthermore, the government has proposed to employers to keep monthly salaries at pre-cut levels, without explicitly legislating. Indeed, over 90% of employees did not have any change in the level of remuneration (Cahuc et al., 2008). Later, in 1996 through the Robien Act, voluntary reductions in working time were introduced by 10-15% and employment increased by 10% in exchange for a reduction of insurance contributions by 7 years. However, unemployment has not declined to a sufficient degree. Thus, in 1998 the Socialist Co-operative Government of Lionel Jospin (1997-2002) launched an ambitious plan to reduce the working time from 39 to 35 hours in order to reduce unemployment. The reduction of working hours this time was compulsory and took into account the economic impact of reduced working time on businesses in contrast to the 1982 legislation. This reduction was made through two legislative initiatives named after Labor Minister Martine Aubry.

Originally adopted in June 1998, the Aubry I Act gave financial incentives (grants) to businesses to reduce the working time of their staff in order to increase the number of people employed. A company had to reduce its working time by at least 10% and increase its number of employees by at least 6% to be eligible for a grant. For each employee whose working hours were reduced or for each new hire, the number of employer contributions was subsidized for five years. The legislation has distinguished between large (more than 20 employees) and smaller businesses, recognizing that smaller ones need more time to adjust to new data. The reduction in working time could take place by 2002 for businesses with fewer than 20 employees. To facilitate this transition, for smaller firms the law reduced the cost of overtime remuneration and increased the overtime ceiling to continue to employ their staff for 39 hours, paying the extra overtime cost. Compared to the "Robien Law" the grants were not based solely on the increase in recruitment but also on the reduction of working time. Because most small businesses did not have statutory labor unions, the legislation was intended to influence the way trade unionism works. In order to benefit from the favorable provisions of the law, businesses employing less than 50 employees would have to sign firm level agreements that the reduced working hours were apparent (Chemin & Wasmer, 2009).

Two years later, in January 2000, the second Aubry Act (called Aubry II) was enacted. This law again expanded the range of grants to businesses in two ways: on the one hand, it subsidized FRF 4,000 per year and per employee reduced the working hours to 35 per week. On the other hand, it subsidized part of the payroll costs of those with low or average remuneration. The amount of grant amounted to FRF 17,500 per employee who was paid the minimum wage (Cahuc et al., 2008). These financial incentives concerned companies that were not otherwise subsidized at that time. The main difference from Aubry I is that there was no minimum number of new jobs that needed to be created for businesses to be eligible for state support. Even after the end of the period in which the employer contributions were subsidized, the payroll cost subsidy continued to exist for those firms that agreed to 35 working hours provided they expressed a desire to create or save jobs (Hayden, 2006). Alongside these measures, a minimum guaranteed wage has been instituted which has been increasing annually on the basis of inflation so as not to reduce the monthly allowance of workers. Subsequently, in January 2003 the Fillon Act reduced the incentives to businesses that had not completed the transition to 35 business hours per week but were not fully abolished. The legislation was relaxed even more in 2005 as it became easier and more cost-effective for businesses to pay their employees for overtime, and employees are entitled to work longer than 220 hours a year provided they do not exceed 44 hours per week on average. Grants were no longer linked to a reduction in working time. The result was a loss to businesses that reduced their employees' working time to 35 hours a week while their competitors stayed 39 hours and received a partial subsidy for employer contributions from the state.

## Evaluation of the French model of reduction of working time

By looking at the French example, we aim to examine whether reducing working hours is a way to counter the potential contraction of employment due to technological change in a competitive market. It is recalled that, according to the predictions of competitive market theory, when labor time is reduced externally, as in the case of a legislated reduction in working time, production costs increase and thus employment is reduced. In the case of France, this happened in the first generation of reforms in 1982. Following the change in legislation, subsidies and tax incentives were provided for businesses to reduce the cost-neutral time, substantially confirming the predictions of competitive market theory. The French case is of particular interest because, over the last decades, the aim of reducing working time has been seeking to increase employment in different ways, each of which has produced different results. Essentially in the same economy, with just a few years of deviation from the same demand, the reduction in working time has had diametrically different effects on employment. The key difference is the rising cost factor as predicted by the competitive market model. As the working hours decrease, the marginal cost of production increases, resulting in a reduction in the optimum amount of product produced. This ,in turn, reduces the demand for labor. An extension, as will be shown by the policy of reducing working hours per se, is not a way of dealing with the possible contraction of employment, especially when this contraction is partly or entirely due to technological change. Higher marginal production costs will cause workers to be replaced by newer capital equipment. Consequently, in a competitive market experiencing the effects of technological change, a reduction in working hours could lead to a reduction in employment even if hourly earnings remain constant.

In order to understand the consequences of the reduction in working time, we will first look at the reduction of working hours in 1982. As mentioned above, this reduction was sudden and unexpected but it took some years for it to be implemented. On the eve of the French elections in 1981, Francois Mitterrand's victory was unlikely, as most polls showed Giscard d'Estaing being re-elected. Similarly, the parliamentary elections took place after the presidential elections, therefore, although the reduction of working time was a question on the socialists' pre-election agenda, however, given the demographic picture, the implementation of the reduction was unexpected. The cut of working time occurred in February 1982 but the measure was implemented very slowly. According to official statistics, in April 1982 when the standard labor force survey was carried out, only a few companies had managed to harmonize the working week of their employees.

From 28% in 1982 the proportion of workers working 39 or 40 hours a week dropped to 20% in 1983-1985 (Crépon & Kramarz, 2002). In order to study the impact of the sharp decline in working hours, Crepon and Kramarz (2002) analyzed labor force surveys for the years 1977-1987 by comparing workers who worked 40 or more hours a week with those who worked less than 39. Their findings showed that workers who worked for 40 hours or more a week in March 1981 were less likely to remain in place after the reform compared to the same characteristics (educational, racial, etc.) workers employed for 36-39 hours weekly. In addition, they concluded that those who continued to work 40 hours in 1982 were more likely to lose their job than those who worked less than 39 hours. All of their findings were statistically significant and showed a negative correlation between the reduction in working time and employment.

Combined with legislative interventions that not only did not allow for a reduction in the proportion of working hours but increased the minimum wage by 5% since July 1981, the abovementioned adverse effects mainly affected those remunerate around the minimum wage level. These findings are confirmed by the model of Abowd et al., (1996), which estimates that this increase in the minimum wage will result in an increased rate of destruction in low-wage jobs (about 8%), which corresponds to about 2% of the rate of job loss in the economy annually. The result is that labor cost elasticity is less than -1, indicating inelastic demand for labor, a consequence of increased wage costs and reduced working time (Crépon & Kramarz, 2002). In conclusion, by looking at the unemployment figures for the period, we see an increase after the reduction of working time. In 1982 unemployment was at 6.6% while in subsequent years it rose to 7.3% in 1983, 8.4% in 1984 and 8.7% in 1985. In the next major reduction in working time (at 35 hours under the Robien and Aubry I laws), although wage levels and other non-wage costs remained steady, corporate grants began to apply. This State aid was intended to absorb the negative effects of employment on the reduction of working time again. Looking at unemployment rates in 1997, before the implementation of Aubry I, we can see that unemployment declined sharply from 10.7% in 1997 to 7.8% in 2001 following the implementation of grants. Similarly, unemployment rose again to 8.5% when tax incentives began to be re-examined in 2003. This increase in unemployment is largely linked to the subsidies given to businesses.

So while the subsidized reduction in working time has actually created jobs, it does not reflect the full picture. During the period 1999-2001, unemployment in France decreased by 2.2% after a decade of persistently high unemployment. Unemployment on the European Union average dropped by 1.2%. Unemployment rate reduction in France is higher but taking into account that the average GDP growth in France at that time was 3.1% while in the European Union 2.5% it is concluded that unemployment is largely related to the more favorable economic environment in France at that time. According to the research of Boeri et al., (2008) between 1997 and 2000, firms that had reduced Aubry I's employment time experienced a 10.5% increase in employment. Of this percentage, the increase attributable to improved demand in the economy (not linked to a reduction in working time) was 5%, while a 2% is attributed to the decrease in labor costs that were linked to the existence of the subsidies. The reduction in working hours contributed to a 3.4% increase in employment. At this rate, of course, the contribution of grants to businesses should be taken into account as through tax exemptions ensure the viability of the business. Without them, they might have reduced their employment or even ceased their operation.

In order to calculate the increase in employment that is linked to the Robien and Aubry laws, the number of jobs created following the implementation of these laws was examined. From this figure, the number of jobs that would have been created without the incentives provided by the state was subtracted. This number was calculated on the basis of the jobs created in similar typical businesses that remained despite the motivation given by the state at 39 hours a week. The result showed that the Robien law increased employment by 7.2% while the Aubry law increased by 7% (Jugnot, 2002 as mentioned in Boisard, 2004). If we take into account however, the financial cost of corporate grants then the way we evaluate the effects of the reduction in working time may also change. In general, the way in which the reduction of working time in France was implemented following the Aubry legislation is cost-neutral for businesses and that was the reason why employment was not negatively affected as it was in 1982. However, it very high the cost of money or tax exchanges given to businesses by the state.

It was estimated that the cost of implementing the program in 2003-4 reached 6 billion euros, creating or saving 350,000 jobs (Estevao & Sa, 2006). However, this amount does not include the costs resulting from the reduction of working time and the corresponding grants for employees in the public sector. According to the most pessimistic calculations, if the public sector is also calculated, the cost can reach 16-22 billion euros. If this number divided by 350,000 jobs that it is estimated that were created or saved by the measure, then the cost-per-job reaches the 45,700-62,800 euros (Askenazy, 2008; Heyer, 2013). Of this amount, we should deduct government revenue from additional personal income taxes due to higher incomes, increased value-added tax due to an increase in consumption (about EUR 3.7 billion) and reduced costs for unemployment benefits (1, EUR 8 billion) (Heyer, 2013). Therefore, the cost of the subsidy measure is estimated at 10.5-16.5 billion euros, ie 30,000-47,000 euros per year per job. Of course we cannot calculate the exact amount of grants because it is correlated with the number of jobs, but in any case, it is a respectable size that, if enlarged, can cause financial problems larger than it was intended to solve.

## Conclusions regarding french working time reduction

As mentioned above, the example of France was used to examine whether reducing working hours is a way of dealing with the possible contraction of employment in a competitive market environment. In the preceding sections, we have shown that the first wave of reduction in working time in 1982 confirms the predictions of the competitive market model. After applying the reduction in working time, there was no positive correlation between unemployment and employment. On the contrary, the unemployment rate and the rate of job destruction have increased in response to the reduction in working hours, especially for wage earners near the bottom. The second wave of reductions in working time came when the working week was reduced to 35 hours from 39. This legislative effort was different from the previous one as it gave large tax and financial incentives to businesses to implement the reduction in working time. The purpose of government grants was to make the reform cost-neutral for businesses. Indeed, this reform increased the number of jobs as it did not affect the unit cost of the product for employers. However, this perspective is incomplete because it does not address the cost of government grants. If we take into account the cost of this measure, which is estimated that may reach EUR 47,000 per year per job, we realize that reducing working time is by no means a sustainable way of tackling rising unemployment.

## MATERIAL AND METHODS

*Working time and employment meta-analysis*

TABLE 2. LISTS OF SURVEYS

| Study | Country | Data | Elasticity |
|---|---|---|---|
| (Hunt, 1999) | W. Germany | Observations from 30 manufacturing industries between 1982 and 1993 | -1,52 for observations from 30 industries (statistical insignificant) <br><br> 0,20 for observations from 10 industries <br><br> *If we do not take into account the increase in hourly wage costs due to the reduction in working time, the elasticity is 0.96 (30 industries)* |
| (Andrews, Schank, & Simmons, 2005) | East and West Germany | 13315 observations in total. For the W. Germany, 4561 in the field of manufacturing and agriculture and 3823 in the services sector. For the E. Germany 2656 observations in the processing and agriculture sector and 2275 in the service sector. | East and West Germany all sectors: -0.06 (statistically insignificant) <br> East Germany (manufacturing and agriculture): -0.75 and -0.17 (services-statistically insignificant) <br> West Germany: 0.08 (manufacturing and agriculture) and 0.06 (services). <br><br> *These elasticities have not taken into account the increase in hourly wage costs due to reduced working hours* |
| (Crépon & Kramarz, 2002) | France | Labor Force Survey. Sample of 22345 individuals between 1977-1987 | 0.256 (employees on wages above the minimum wage). <br><br> 0,8-1,6 (if the impact of working time reduction on wage costs is not taken into account) |
| (Kapteyn, Kalwij, & Zaidi, 2004) | 16 OECD countries | Annual time series. 493 observations. | -0,16 (statistically insignificant) |
| (Skuterud, 2007) | Canada | Annual time series 1996-2002. 254 observations | 0,05 (male) /-0,20 (female) |
| (Hart, 1987) | W. Germany | Observations on manufacturing industries 1969-1981 | -0,39 ( statistically insignificant ) |
| (Brunello, 1989) | Japan | Monthly time series of the manufacturing industry 1973-1986. 7.6 million observations | -0,0216 ( statistically insignificant ) |
| (Steiner & Peters, 2003) | W. Germany | Annual observations from 27 manufacturing industries during the period 1978-1995 | statistically insignificant for all categories. <br> 0,43 (unskilled workers) <br> 0,19 (medium skilled workers) <br> 0 (high skilled workers). |
| (Estevao & Sa, 2008) | France | Annual time series of years 1999-2002 <br> Sample of 10286 men and 5662 women | 0,22 (male)/ -0,00024 (female) <br><br> *These elasticities have not taken into account the increase in hourly wage costs due to reduced working hours* |

This section will examine the existing empirical studies on the impact of reducing working time in order to draw more general conclusions about whether reducing working hours is a way to deal with the possible contraction of employment (Table 2). In order to achieve this objective, 9 empirical studies were examined, of which 18 observations / elasticities of the working time were extracted. Of these 18 observations, 9 (50%) were statistically insignificant, 6 (33.33%) indicated that a decrease in working time caused a decrease in employment, and 3 (16.67%) indicated that a decrease in working time increased employment. However, several of the observations (5 in number - 27.78%), although statistically significant, had very little difference from zero.

Some of the studies listed in Table 2 do not weigh their findings with regard to the increase in remuneration as a direct or indirect consequence of the reduction in working time, so in the subsequent analysis, they will not be taken into account to show the impact of the reduction of working time ceteris paribus.

The first research to be examined is by Hunt (1999). This research examines the case of Germany where the reduction in working time was achieved through an agreement to maintain a fixed amount of monthly earnings received before the reduction, increasing the hourly labor costs. Looking at a sample of 30 industry sectors over the period 1984-1994, one concludes that a one hour reduction in working time (from 40 to 39 hours - 2.5%) if the employment reduction resulting from increased wage costs is not weighed, reduces employment by 2.4%. When the findings are weighted to account for the increase in hourly labor costs, it concludes that a 2.5% reduction in working time reduces employment by 0.5% (when examining a sample of 10 industries) and 3.8% (which however has a large standard error and is not statistically significant) when examining 30 industries.

The second empirical study (Andrews et al., 2005) has not been included in the second meta-analysis because as it does not take into account the impact of reduced working time on shaping hourly labor costs. It deals with companies engaged in the processing, agriculture and service sectors of West and East Germany from 1993-1999. The findings show a large difference in the flexibility of employment in reducing the working time between East and West Germany largely due to different economic structures. In East Germany, a 4 hour (10.26%) reduction in working time greatly increased employment (up to 7.74%), while in the case of West Germany employment remained virtually unaffected (decreased by 0,82%).

The third survey (Crépon & Kramarz, 2002) was conducted in France after the reduction of working time and the increase in the minimum wage in 1982. This study examined the impact of reducing working time through the possibility of an employee losing his or her job. The authors estimate that a 2.5% reduction in working time causes a 2-4% reduction in employment when the increase in wage costs is not taken into account. In order to take into account the increase resulting from both a 5% increase in the minimum wage and a statutory provision banning a reduction in the monthly wage proportional to work, they examined the likelihood of unemployment for the period 1982-1984 for which there were statistics on wage costs. According to these calculations, a 2.5% reduction in working time resulted in a 0.64% decrease in employment for workers who did not receive a pay rise. Those paid with the minimum wage took a 5% pay rise. As a result, employment was reduced by 8,34%. The decline in employment first hit workers with the minimum wage, and after 2 years the rest of the workforce was affected by the reduction in working time.

Subsequently, the Skuteruds' (2007) survey in Canada between 1997 and 2000 examined the reduction of working time in Quebec from 44 to 40 hours per week. The peculiarity of this research is that, unlike those who calculated the change in working time in Europe, in the case of Quebec there was no restriction on maintaining the level of the monthly wage, while the measure concerned workers who were not members of labor unions, they were relatively unskilled and had high rates of unemployment. Consequently, all of the above features plus the fact that employers did not compensate employees for the reduced hours, making this case ideal for studying the reduction of working time as a job creation strategy. This study showed that a 9.09% reduction in working time failed to substantially increase employment. Specifically, when calculating the average of all sectors of the economy, the decline in working time by 9.09% causes a 0.5% decrease in employment for men and a 2% increase for women.

Finally, the last research showing statistically significant observations is that of Estevao and Sa (2008). Its methodology is similar to that of Crepon and Kramarz. It essentially compares businesses over 20 people with smaller businesses that were not immediately forced to apply the cuts and calculates the likelihood of an employee becoming unemployed in each case. This study does not weigh employment outcomes based on the increase in hourly costs. However, it appears a tendency of an increase in the number of employees who work in two different companies by 0.7% after the law was implemented, while the employment decline ranged between 1 and 3.9% (decreasing as the years passed from 1999 to 2002). In conclusion, the authors observed that although the essence of the law was to increase employment, no such increase was observed in large enterprises over smaller ones.

In order to give a more precise picture of the issue, we will proceed with an analysis of the aforementioned empirical studies. First, we will perform the meta-analysis by excluding only the elasticities that are not statistically significant (Table 3a). We will then exclude, except for statistically insignificant elasticities and those which have been derived from calculations that do not weigh the results of empirical studies on the basis of the increase in hourly costs (Table 4). In this way, we will have a clearer picture of the impact of working time reduction.

TABLE 3A. TABLE WITH STATISTICAL SIGNIFICANT ELASTICITIES

|                  | N     | r        | NxR       |
|------------------|-------|----------|-----------|
| Hunt             | 30    | 0,96     | 28,8      |
| Andrews et al.   | 2656  | -0,75    | -1992     |
| Andrews et al.   | 4561  | 0,08     | 364,88    |
| Andrews et al.   | 3823  | 0,06     | 229,38    |
| Crepon/Kramarz   | 22345 | 1,2      | 26814     |
| Skuterud         | 254   | 0,05     | 12,7      |
| Skuterud         | 254   | -0,2     | -50,8     |
| Estevao/ Sa      | 10286 | 0,22     | 2262,92   |
| Estevao/ Sa      | 5662  | -0,00024 | -1,35888  |
|                  |       |          |           |
| TOTAL:           | 49871 |          | 27668,5211|

$$r = \frac{NxR}{N} \quad r = \frac{27668.5211}{49871} \quad r = 0.5548$$

In this way, by using the data in Table 3a we calculate the weighted effect size. In order to calculate confidence intervals, we created a new table (Table 3b).

TABLE 3B. TABLE USED TO CALCULATE THE VARIANCE OF EFFECT SIZE AND THE SAMPLING ERROR

|                  | N     | r        | NxR       | n( r- (0,5548))² |
|------------------|-------|----------|-----------|------------------|
| Hunt             | 30    | 0,96     | 28,8      | 4,9256           |
| Andrews et al.   | 2656  | -0,75    | -1992     | 4521,848         |
| Andrews et al.   | 4561  | 0,08     | 364,88    | 8,9849           |
| Andrews et al.   | 3823  | 0,06     | 229,38    | 935,9738         |
| Crepon/Kramarz   | 22345 | 1,2      | 26814     | 9301,8445        |
| Skuterud         | 254   | 0,05     | 12,7      | 10,0183          |
| Skuterud         | 254   | -0,2     | -50,8     | 51,1155          |
| Estevao/ Sa      | 10286 | 0,22     | 2262,92   | 1152,9684        |
| Estevao/ Sa      | 5662  | -0,00024 | -1,35888  | 1744,2889        |
|                  |       |          |           |                  |
| TOTAL:           | 49871 |          | 27668,5211| 17731,9679       |

Based on the second and fifth columns, the variance of effect size and the sampling error will be calculated to correct the effect size variation based on these measurements.

$$\sigma_r^2 = \frac{17731.9679}{49871} \qquad \sigma_r^2 = 0.3555$$

The sampling error can also be calculated directly from the data in the table. The only size not readily available is the average sample size. To calculate this, we will divide the total calculated in the second column by the number of studies we are considering.

$$\hat{\sigma}_e^2 = \frac{(1-\bar{r}^2)^2}{\bar{N}-1} \qquad \sigma_e^2 = \frac{0.47913}{5540.2222} \qquad \sigma_e^2 = 0,0000864821$$

Having calculated these two sizes, we can now calculate the variation of the effect size of our sample.

$$\sigma_\rho^2 = \sigma_r^2 - \sigma_e^2 \qquad \sigma_\rho^2 = 0.3555 - 0.0000864821 \qquad \sigma_\rho^2 = 0.3554$$

With this number, we calculate the confidence interval (95%).

$CI_{upper}$ = r + 1,96$\sqrt{\sigma_\rho^2}$ = 1.7233     $CI_{lower}$ = r − 1,96$\sqrt{\sigma_\rho^2}$ = -0.6137

Average elasticity: 0.5548

Based on the above, we can conclude that a preliminary 1% reduction in working time will result in a 0.5548% reduction in employment. However, the impact of reducing working time on the cost of working time has not been taken into account in this measurement. Even if monthly salaries remain unchanged as an absolute number, reducing working time increases the hourly labor costs. If we follow the aforementioned logic by calculating only those elasticities that are weighted to account for the increase in hourly costs, we will have a fuller picture of how the reduction in working time affects employment when employees are commensurate with their working time. The empirical studies that have made such a weighting are significantly less as shown in Table 4a.

TABLE 4A. EMPIRICAL STUDIES WEIGHTED TO ACCOUNT FOR THE INCREASE IN HOURLY COSTS

| | N | R | NxR |
|---|---|---|---|
| Hunt | 10 | 0,2 | 2 |
| Crepon/Kramarz | 22345 | 0,256 | 5720,32 |
| Skuterud | 254 | 0,05 | 12,7 |
| Skuterud | 254 | -0,2 | -50,8 |
| | | | |
| TOTAL | 22863 | | 5684,22 |

$$r = \frac{NxR}{N} \quad r = \frac{5684.22}{22863} \quad r = 0.2486$$

TABLE 4B. TABLE USED TO CALCULATE THE VARIANCE OF EFFECT SIZE AND THE SAMPLING ERROR

| | N | R | NxR | n( r- (0,2486))² |
|---|---|---|---|---|
| Hunt | 10 | 0,2 | 2 | 0,0236 |
| Crepon/Kramarz | 22345 | 0,256 | 5720,32 | 1,2236 |
| Skuterud | 254 | 0,05 | 12,7 | 10,0183 |
| Skuterud | 254 | -0,2 | -50,8 | 51,1155 |
| | | | | |
| TOTAL | 22863 | | 5684,22 | 62,381 |

$\sigma_r^2 = \frac{62,381}{22863}$     $\sigma_r^2 = 0,0027$

$\hat{\sigma}_e^2 = \frac{(1-\bar{r}^2)^2}{\bar{N}-1}$     $\sigma_e^2 = \frac{0.88021}{5714.75}$     $\sigma_e^2 = 0.0001540$

$\sigma_\rho^2 = \sigma_r^2 - \sigma_e^2$     $\sigma_\rho^2 = 0.0027 - 0.0001540$     $\sigma_\rho^2 = 0.002546$

$CI_{upper}$ = r + 1,96$\sqrt{\sigma_\rho^2}$ = 0.3475     $CI_{lower}$ = r − 1,96$\sqrt{\sigma_\rho^2}$ = 0,1497

Average elasticity: 0.2486

## Discussion

Based on the above findings we conclude that when the increase in hourly labor costs is weighted, employment is expected to decrease by 0.25% for every 1% decrease in working time. Compared to measures that do not weigh labor costs growth (elasticity 0.55) we find that the impact on employment in both cases remains negative, but the employment decline is dramatically higher when workers' salaries are not proportionally reduced. Given the number of empirical surveys that are statistically insignificant or show resilience very close to zero, it is not easy to form a clear picture of the relationship between employment and working time. Although the elasticity we calculated through the meta-analysis shows that in any case, the reduction of working time does not increase employment, we should consider how the reduction of working time is decided. Empirically observing the method of application, we find that in most cases the labor unions press on and eventually achieve steady wage earnings despite the reduction in working time. Consequently, a large increase in hourly labor costs is an intrinsic part of reducing the working week. With this in mind, it may be that the real impact of the reduction in working time is closer to the elasticity calculated in the first case where the increase in hourly labor costs was not weighted. However, we also proceeded with the second meta-analysis to consider the more favorable scenario, namely a reduction in working time with a corresponding reduction in earnings. Even under these conditions, it seems that reducing working time does not improve employment and it is doubtful the reduction of working time to evolve in a way of dealing with the possible contraction of employment.

## CONCLUSIONS

The purpose of this article was to explore whether reducing working time could be a solution to the problem of rising and persistent unemployment that plagues many economies. In order to achieve this objective, models of the competitive market, of collective bargaining and of monopsony market were examined. The competitive market model predicts a decline in employment when working time is reduced as a result of the increase in hourly production costs. The collective bargaining model predicts that the impact on employment is determined, inter alia, by the strength of the bargaining partners and the economic environment. It is likely in that case that employment growth could be due to increased productivity and extended use of capital equipment. Finally, in a monopsony market a slight reduction in working hours would temporarily improve employment. However, in the long run, there would be no significant impact on stabilizing the economy at higher employment levels. Subsequently, the case studies of the Netherlands and France were examined, where these two countries are effectively used as a means of evaluating in practice the models of the competitive market and collective bargaining. In both cases, employment outcomes were not as expected as they did not demonstrate a radical and long-term solution to the problem of unemployment.

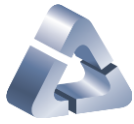

**Journal of Applied Economics and Business**

In conclusion, in the last section, we examined empirical research on the impact of reducing working time on employment. Our analysis presumes that reducing working time reduces employment. Specifically, a 1% reduction in working time causes a 0.25% - 0.55% reduction in employment. Based on the abovementioned we conclude that reducing working time is not a way to reduce unemployment and increase employment by substantially confirming the theoretical predictions of the competitive market model. The reduction of working time ultimately leads to an increase in hourly costs which affects both the unit cost of the product and the competitiveness of business and economy in general, without providing employment benefits.

## ACKNOWLEDGEMENT

This research is co-financed by Greece and the European Union (European Social Fund - ESF) through the Operational Programme "Human Resources Development, Education and Lifelong Learning" in the context of the project "Strengthening Human Resources Research Potential via Doctorate Research" (MIS-5000432), implemented by the State Scholarships Foundation (IKΥ).

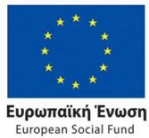
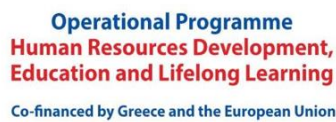
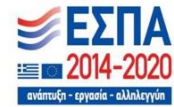